\documentclass[aps,prl,reprint,amsmath,amssymb,showpacs,nobibnotes]{revtex4-1}

\usepackage{graphicx}
\usepackage{bm}     
\usepackage{hyperref} 
\usepackage{dsfont}    
\usepackage{color}
\usepackage[normalem]{ulem}
\usepackage{cleveref}

\newcommand{\ket}[1]{|#1\rangle}
\newcommand{\braket}[2]{\langle{#1}|{#2}\rangle}

\newcommand{\bra}[1]{\langle#1|}
\newcommand{\proj}[1]{\ket{#1}\!\bra{#1}}
\newcommand{\proju}[2]{\ket{#1}\!\bra{#2}}

\newcommand{\ab}[4]{\ket{{#1}_H,{#2}_V}_A \ket{{#3}_H,{#4}_V}_B}

\newcommand{\eq}[1]{Eq.~(\ref{#1})}
\newcommand{\fig}[1]{Fig.~\ref{#1}}

\def\beq{\begin{eqnarray}}
\def\eeq{\end{eqnarray}}

\newcommand{\single}{\text{s}}
\newcommand{\three}{\text{t}}
\newcommand{\expec}[2]{\langle#1#2\rangle}

\begin{document}

\title{Remote preparation of three-photon entangled states via single-photon measurement}

\author{Young-Sik Ra}
\altaffiliation{Present address: Laboratoire Kastler Brossel, Universit\'e Pierre et Marie Curie, 75252 Paris, France}
\email{youngsikra@gmail.com}
\affiliation{Department of Physics, Pohang University of Science and Technology (POSTECH), Pohang, 790-784, Korea}

\author{Hyang-Tag Lim}
\altaffiliation{Present address: Institute of Quantum Electronics, ETH Zurich, CH-8093 Zurich, Switzerland}
\affiliation{Department of Physics, Pohang University of Science and Technology (POSTECH), Pohang, 790-784, Korea}

\author{Yoon-Ho Kim}
\email{yoonho72@gmail.com}
\affiliation{Department of Physics, Pohang University of Science and Technology (POSTECH), Pohang, 790-784, Korea}

\date{\today}

\begin{abstract} 
Remote state preparation (RSP) provides an indirect way of manipulating quantum information based on the nonlocal effect of quantum measurement. Although RSP has been demonstrated in recent years to remotely prepare multi-photon states, quantum measurement on the same number of photons was required, i.e., to prepare $N$-photon states via RSP, quantum measurement on the other N-photons was required, hence significantly limiting practicality and applicability of RSP. Here we report the first experimental demonstration of remote preparation of three-photon entangled states by measuring only a single-photon entangled with the three photons. We further generalize our protocol to prepare multi-photon entangled states with arbitrary photon number and purity via single-photon measurement. As our RSP scheme relies on the nonlinearity induced by single-photon measurement, it enables quantum state engineering of multi-photon entangled states beyond the linear optical limit. Our results are expected to have significant impacts on quantum metrology and quantum information processing. 
\end{abstract}
%

\maketitle


Entanglement, once conceived as the `weirdness' of quantum mechanics~\cite{Einstein:1935aa}, is now at the heart of quantum technologies, such as, quantum communication~\cite{Briegel:1998aa,Kimble:2008aa}, quantum computing~\cite{Knill:2001aa,Raussendorf:2001aa}, and quantum metrology~\cite{Giovannetti:2004aa,Escher:2011aa}. Manipulation of quantum information based on the nonlocal nature of quantum measurement on entangled quantum systems can be categorized into two: quantum teleportation~\cite{Bennett:1993aa,Bouwmeester97,Kim:2001aa} and remote state preparation (RSP)~\cite{Pati:2000bb,Bennett:2001aa}. In quantum teleportation, Alice can transfer an \emph{unknown} qubit to Bob by sending him two bits of classical information resulting from the Bell state measurement \cite{Kim:2001aa}, assuming Alice and Bob already share a pair of maximally-entangled qubits.  In RSP, Alice and Bob are still assumed to share a pair of maximally-entangled qubits, but Alice can transfer a \emph{known} qubit to Bob by sending him only one bit of information from her measurement. Moreover, as RSP does not require complete Bell state measurement, it can be more readily scaled up for larger quantum systems.

In photonic systems, RSP has recently been demonstrated for single-photons \cite{Peters:2005aa,Barreiro:2010aa,Solis-Prosser:2011aa}, two photons~\cite{Mikami:2007aa,Ra:2015bb}, and three photons~\cite{Radmark:2013be}. In these RSP schemes, however, quantum measurement on the same number of photons was required, i.e., to prepare an $N$-photon state at Bob, $N$-photon quantum measurement is required at Alice, hence significantly limiting practicality and applicability of RSP. However, RSP does not necessarily require Alice to measure the same number of photons as the remotely prepared quantum state on Bob. In fact, if Alice and Bob share an entangled state consisting of a single-photon and multiple photons, single-photon measurement at Alice would be sufficient to remotely prepare a multi-photon state at Bob.

In this work, we demonstrate remote preparation of various three-photon entangled states via single-photon measurement by preparing entanglement between a single-photon and three single-photons. Alice's measurement on the single-photon remotely prepares a three-photon entangled state in the form of $\alpha \ket{2,1} + \beta e^{i\theta} \ket{1,2}$ at Bob. We also generalize our RSP protocol to prepare multi-photon entangled states of arbitrary photon number and purity via single-photon measurement. Our RSP protocol extends the capability of multi-photon state engineering beyond the linear optical limit via the nonlinearity induced by the single-photon measurement~\cite{Knill:2001aa,Lanyon:2008ci}, allowing us to prepare various multi-photon states required for quantum metrology ~\cite{Huver:2008aa,Demkowicz-Dobrzanski:2009aa,Matthews:2011aa,Tichy:2014aa} and for fundamental studies in quantum optics ~\cite{Sehat:2005ig, Bjork:2012aa}.


Remote preparation of a three-photon entangled state by single-photon measurement requires Alice and Bob to share an entangled state between a single-photon and three single-photons of the form,
\beq \ket{\Phi}_{AB} = \frac{\ket{1_H,0_V}_A \ket{1_H,2_V}_B + \ket{0_H,1_V}_A \ket{2_H,1_V}_B}{\sqrt{2}},~~~~  \label{eq:entangledstate} \eeq
where subscripts $A$, $B$, $H$, and $V$ refer to Alice, Bob, horizontal polarization, and vertical polarization, respectively. Alice measures her single-photon in the basis $\{ \ket{\phi}_A,\ket{\phi_\bot}_A \}$, where $\ket{\phi}_A=\beta \ket{1_H,0_V}_A + \alpha e^{i \theta} \ket{0_H,1_V}_A$ with real parameters $\alpha, \beta$, and $\theta$. Note that $_A\braket{\phi_\bot}{\phi}_A=0$. When Alice's single-photon is measured in the basis $\ket{\phi}_A$, Bob's three-photon state is then projected onto
\beq \ket{\psi}_{B} =  \alpha \ket{2_H,1_V}_B + \beta e^{i\theta} \ket{1_H,2_V}_B . \label{eq:three-photon} \eeq
Therefore, the single-photon measurement at Alice can remotely prepare the desired three-photon entangled state at Bob.

\begin{figure*}[t]
\centerline{\includegraphics[width=12cm]{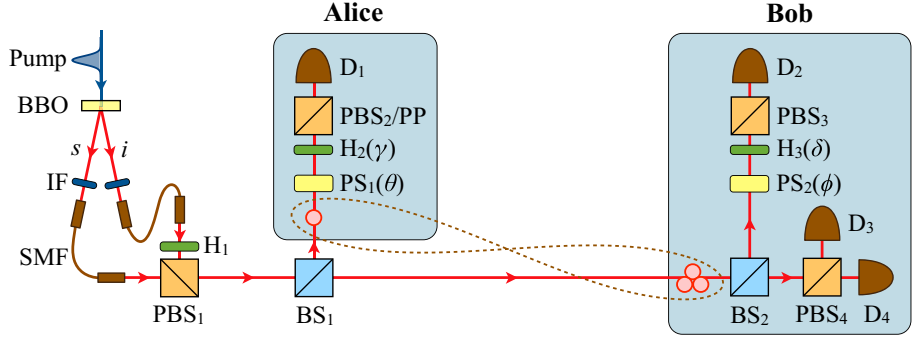}}
\caption{\textbf{Experimental setup.} 
At the output of a polarizing beam splitter PBS$_1$, the quantum state of four photons is $\ket{2_H,2_V}$. When a single-photon is reflected and the other three photons are reflected at a non-polarizing beam splitter BS$_1$, Alice and Bob share the entangled state between the single-photon and the three photons in \eq{eq:entangledstate}. Alice measures the single-photon by projection onto $\cos2\gamma \ket{1_H,0_V}_A + e^{i\theta}\sin2\gamma \ket{0_H,1_V}_A$ using a phase shifter PS$_1(\theta)$, a half wave plate H$_2(\gamma)$, PBS$_2$, and a single-photon detector D$_1$. As a consequence, the three-photon entangled state in \eq{eq:three-photon} is prepared at Bob, and he measures the three photons by projection onto $\cos2\delta\ket{2_H,1_V}_{B}+e^{i \phi}\sin2\delta \ket{1_H,2_V}_{B}$ using PS$_2(\phi)$, H$_3(\delta)$, PBS$_3$, PBS$_4$, and coincidence detection on D$_2$, D$_3$, and D$_4$. 
RSP of a  three-photon mixed state (\eq{eq:mixed}) requires the use of  a partial polarizer (PP)\cite{Peters:2005aa}. BBO, $\beta$-BaB$_2$O$_4$ crystal; $s$, signal; $i$, idler; IF, interference filter; SMF, single-mode fiber. A more detailed explanation of the setup is provided in Methods.
}\label{fig:setup}
\end{figure*}

To prepare entanglement between a single-photon and three photons, we employ the experimental setup shown in \fig{fig:setup}. Four photons in $\ket{2_H, 2_V}$ impinge on a non-polarizing beam splitter BS$_1$, and when a single-photon is reflected to Alice and the other three photons are directed to Bob (assured by detecting the corresponding numbers of photons at Alice and Bob), the single-photon and the three photons are in the entangled state in \eq{eq:entangledstate}. To verify entanglement between the single-photon and the three photons, we test the Clauser-Horne-Shimony-Holt (CHSH) inequality~\cite{Clauser:1969vv}. The Hilbert space for the quantum state under the test has the dimension 2$\otimes$2 because, for the single-photon, the basis is $\{ \ket{1_H,0_V},\ket{0_H,1_V} \}$ and, for the three photons, the basis is $\{ \ket{2_H,1_V},\ket{1_H,2_V} \}$. The CHSH parameter is then defined as
\beq
S_\textrm{CHSH} = |-\expec{\hat{\mu}^\single}{\hat{\mu}^\three} + \expec{\hat{\mu}^\single}{\hat{\pi}^\three} + \expec{\hat{\pi}^\single}{\hat{\mu}^\three} + \expec{\hat{\pi}^\single}{\hat{\pi}^\three}|,
\label{eq:CHSH}
\eeq
 where $\hat{\mu}^\single=\frac{1}{\sqrt{2}} (\hat{\sigma}^\single_z+\hat{\sigma}^\single_x)$ and $\hat{\pi}^\single=\frac{1}{\sqrt{2}} (\hat{\sigma}^\single_z-\hat{\sigma}^\single_x)$ are measurements on the single-photon, and $\hat{\mu}^\three=\hat{\sigma}^\three_z$ and $\hat{\pi}^\three=\hat{\sigma}^\three_x$ are measurements on the three photons. The relevant  operators are defined as,
\beq
\hat{\sigma}^\single_z &=& \proj{1_H,0_V}-\proj{0_H,1_V}, \nonumber \\
\hat{\sigma}^\single_x &=& \proju{1_H,0_V}{0_H,1_V}+\proju{0_H,1_V}{1_H,0_V}, \nonumber \\
\hat{\sigma}^\three_z &=& \proj{2_H,1_V}-\proj{1_H,2_V}, \nonumber \\
\hat{\sigma}^\three_x &=& \proju{2_H,1_V}{1_H,2_V}+\proju{1_H,2_V}{2_H,1_V}.
\label{eq:sigma}
\eeq
In experiment, $S_\textrm{CHSH}$ violates the classical limit of 2 by more than 7 standard deviations, summarized in Table~\ref{tab:CHSH}. See Methods for details of measurements to obtain $S_\textrm{CHSH}$.

\begin{table}[b]
\begin{tabular}{cccc|c}
~~$\expec{\hat{\mu}^\single}{\hat{\mu}^\three}$~~&~~$\expec{\hat{\mu}^\single}{\hat{\pi}^\three}$~~&~~$\expec{\hat{\pi}^\single}{\hat{\mu}^\three}$~~&~~$\expec{\hat{\pi}^\single}{\hat{\pi}^\three}$~~~~&~~~$S_\textrm{CHSH}$~~~\\
\hline
$-$0.69(5)&~0.67(5)&~0.64(5)&0.71(4)&~~2.71(9)\\
\end{tabular}
\caption{
\textbf{Test of CHSH inequality between the single-photon and the three photons.} Experimentally measured normalized correlation values and the CHSH parameter $S_\textrm{CHSH}$. The measurement basis for the single-photon is $\{ \ket{1_H,0_V},\ket{0_H,1_V} \}$  and that for the three photons is  $\{ \ket{2_H,1_V},\ket{1_H,2_V} \}$. Thus, the Hilbert space dimension for the state under the test is $2\otimes2$. The experimental $S_\textrm{CHSH}$ value violates the classical limit of 2 by more than 7 standard deviations, a clear indication that the single-photon and the three photons are entangled. Errors represent one standard deviation. 
}\label{tab:CHSH}
\end{table} 

\begin{figure*}[t]
\centerline{\includegraphics[width=\textwidth]{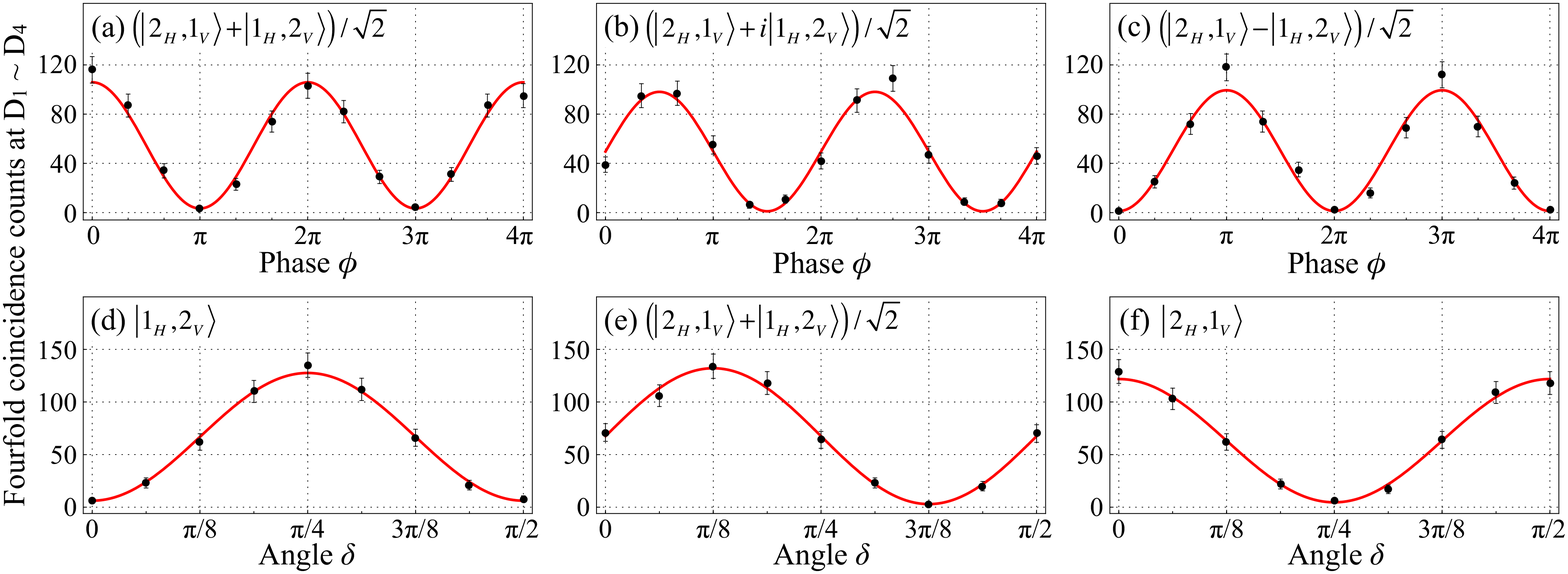}} 
\caption{\textbf{Measurement of remotely prepared three-photon  states.} (a-c) Remotely prepared three-photon entangled states having different phases, in \eq{eq:three-photon_phase}, are measured by projection onto $\frac{1}{\sqrt{2}}(\ket{2_H,1_V}_{B}+e^{i \phi}\ket{1_H,2_V}_{B})$. 
(d-f) Remotely prepared three-photon states having different amplitudes, in \eq{eq:three-photon_amp}, are measured by projection onto $\cos2\delta\ket{2_H,1_V}_{B}+\sin2\delta\ket{1_H,2_V}_{B}$.
Black dots are experimental data, and red solid lines are sinusoidal fittings to the experimental data. Error bars represent one standard deviation by assuming Poissonian counting statistics. Visibilities of the fittings are (a) $93.8\pm1.7\%$, (b) $97.8\pm3.1\%$, (c) $97.4\pm1.6\%$, (d) $90.8\pm1.3\%$, (e) $95.7\pm1.0\%$, and (f) $92.7\pm1.9\%$.
}\label{fig:data}
\end{figure*}

Using the entanglement between the single-photon and the three photons, we carry out RSP of a three-photon entangled-state with an adjustable phase $\theta$, i.e.,
\beq \ket{\psi}_{B}^\text{phase} = \frac{1}{\sqrt{2}}\left(\ket{2_H,1_V}_B + e^{i\theta}  \ket{1_H,2_V}_B \right), \label{eq:three-photon_phase} \eeq
which is useful in quantum metrology with lossy channels~\cite{Huver:2008aa,Demkowicz-Dobrzanski:2009aa,Matthews:2011aa} and in characterizing the kinds of quantum decoherence~\cite{Tichy:2014aa}.
To do this, we employ the single-photon measurement at Alice in \fig{fig:setup}, where the angle $\gamma$ of a half wave plate H$_2$ is set to $\pi/8$, the phase shift $\theta$ of PS$_1$ is adjusted. A click on D$_1$ heralds the successful preparation of the three-photon entangled state in \eq{eq:three-photon_phase}.
For measuring the phase $\theta$ in the remotely prepared three-photon state, we employ the measurement setup at Bob in \fig{fig:setup}, which projects the three-photon state onto $\frac{1}{\sqrt{2}}(\ket{2_H,1_V}_{B}+e^{i \phi}\ket{1_H,2_V}_{B})$; the phase $\phi$ of PS$_2$ is varied while the angle $\delta$ of H$_3$ is set to $\pi/8$, and coincidence counts on D$_2$, D$_3$, and D$_4$ are recorded. By this measurement, the projection probability for the state $\ket{\psi}_{B}^\text{phase}$ is $\frac{1}{2}\left( 1 + \cos(\phi-\theta) \right)$, where the offset of sinusoidal probability as a function of $\phi$ indicates the phase $\theta$ in the three-photon state.
Figures \ref{fig:data}(a-c) show three-photon entangled states having various phases by choosing different $\theta$ values at Alice: for $\theta=0$, the offset of the sinusoidal oscillation is 0, but as $\theta$ is adjusted, the offset is shifted accordingly.

\begin{figure}[b]
\centerline{\includegraphics[width=8.6cm]{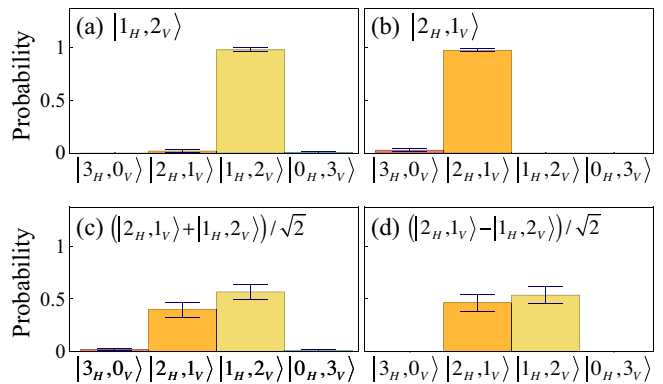}} 
\caption{\textbf{Contributions of $\ket{3_H,0_V}$, $\ket{2_H,1_V}$, $\ket{1_H,2_V}$, and $\ket{0_H,3_V}$ states in experimentally generated three-photon states.} The maximum contribution from $\ket{3_H,0_V}$ and $\ket{0_H,3_V}$ is below 0.03. This result also justifies the choice the Hilbert space dimension $2\otimes2$ as contributions from $\ket{3_H,0_V}$ and  $\ket{0_H,3_V}$ are negligible. Error bars represent one standard deviation.
}\label{fig:contributions}
\end{figure}

We next carry out RSP of a three-photon state at Bob having  a varying degree of entanglement, i.e., 
\beq \ket{\psi}_{B}^\text{amp} = \sin2\gamma \ket{2_H,1_V}_B + \cos2\gamma \ket{1_H,2_V}_B. \label{eq:three-photon_amp} \eeq
To carry out this RSP, the measurement setup at Alice is modified: PS$_1$ is removed, the angle $\gamma$ of H$_2$ is adjusted, and a click at D$_1$ is informed to Bob. At Bob's side, as it is necessary to observe the change of amplitudes in a three-photon state, the projection basis  $\cos2\delta\ket{2_H,1_V}_{B}+\sin2\delta\ket{1_H,2_V}_{B}$ is used. To do so, PS$_2$ is removed and coincidence counts at D$_2$, D$_3$, and D$_4$ are measured as a function of the angle $\delta$ of H$_3$. The projection probability of the state  $\ket{\psi}_{B}^\text{amp}$ by this measurement is $\frac{1}{2}\left( 1 - \cos4(\delta+\gamma) \right)$, and thus, the amplitude in the three-photon state can be obtained by finding the maximum probability, which takes place at $\delta=\pi/4 - \gamma$ (modulo $\pi/2$). Experimental results in \fig{fig:data}(d-f) show the changes of the amplitudes in the three-photon state depending on the single-photon measurement at Alice. Additionally, we have confirmed that contributions of $\ket{3_H,0_V}_{B}$ and $\ket{0_H,3_V}_{B}$ in the generated states are negligible, as shown in \fig{fig:contributions}.

It is important to emphasize that the set of three-photon states in \eq{eq:three-photon_amp} cannot be fully prepared by linear optical transformation of any particular three-photon states. For example, any nontrivial linear optical transformations on $\ket{2_H,1_V}_B$ necessarily generate undesired quantum states $\ket{3_H,0_V}_B$ and/or $\ket{0_H,3_V}_B$, so that the state $1/\sqrt{2} ( \ket{2_H,1_V}_B+\ket{1_H,2_V}_B )$ cannot be prepared. On the other hand, as RSP is based on measurement-induced nonlinearity~\cite{Knill:2001aa,Lanyon:2008ci}, it enables us to access  multi-photon states on different orbits determined by linear optical transformations~\cite{Sehat:2005ig}. Our RSP protocol, thus, extends the capability of multi-photon state engineering beyond the linear optical limit via the nonlinearity induced by the single-photon measurement~\cite{Knill:2001aa,Lanyon:2008ci}, allowing us to prepare various multi-photon states required for quantum metrology ~\cite{Huver:2008aa,Demkowicz-Dobrzanski:2009aa,Matthews:2011aa,Tichy:2014aa} and for fundamental studies in quantum optics ~\cite{Sehat:2005ig, Bjork:2012aa}.


We have so far experimentally demonstrated RSP of pure three-photon states, but the scheme can be further generalized to prepare general multi-photon entangled states of arbitrary purity and  photon number via single-photon measurement.
To prepare a mixed three-photon state at Bob, Alice uses a partial polarizer PP~\cite{Peters:2005aa} instead of PBS$_2$ in \fig{fig:setup}. This then implements a partial projection on the single-photon at Alice of the form,
\beq \mathcal{P}(\ket{\phi}_A\bra{\phi},p) = p \ket{\phi}_A\bra{\phi} + (1-p)I_A /2, \label{eq:partialpol} \eeq
where $p$ is the strength of the projection and $I_A$ is the identity density operator $\ket{1_H,0_V}_A\bra{1_H,0_V}$ $+\ket{0_H,1_V}_A\bra{0_H,1_V}$. Consequently, the three-photon state at Bob becomes a mixed state of the following form,
\beq \rho_B &=& \frac{\text{Tr}_A\left( \mathcal{P}(\ket{\phi}_A\bra{\phi},p)~\ket{\Phi}_{AB}\bra{\Phi} \right)}{\text{Tr}_{AB}\left( \mathcal{P}(\ket{\phi}_A\bra{\phi},p)~\ket{\Phi}_{AB}\bra{\Phi} \right)} \nonumber \\
 &=& p \ket{\psi}_B\bra{\psi} + (1-p) I_B /2, \label{eq:mixed} \eeq
where $I_B=\ket{2_H,1_V}_B \bra{2_H,1_V}$ $+\ket{1_H,2_V}_B\bra{1_H,2_V}$. In other words, an arbitrary amount of noise ($1-p$) can be added to any three-photon pure state $\ket{\psi}_B$ via Alice's partial projection measurement. Note that a pure three-photon state cannot be transformed to such a mixed state by applying the distinguishability-based decoherence schemes~\cite{Kwiat:2000aa,Tichy:2014aa}, which is commonly adopted to generate a mixed single-photon state. This is due to the fact that the distinguishability-based decoherence schemes cause emergence of non-trivial multi-photon state components~\cite{Ra:2013aa,Tichy:2014aa,Tillmann:2015bq}; e.g., introducing $\ket{2_H,\tilde{1_V}}_B$, $\ket{1_H,1_V,\tilde{1_V}}_B$, and $\ket{1_H,\tilde{2_V}}_B$, where $\ket{\tilde{n_V}}$ represents the vertically polarized $n$-photon state in a physically distinguishable mode (i.e., time, path, etc) from that of $\ket{n_V}$.

Now, for preparing a multi-photon state with arbitrary photon number, we consider $n$-pair photon generation in the spontaneous parametric down-conversion. The quantum state at the  output of PBS$_1$ in \fig{fig:setup} is then given as $\ket{n_H,n_V}$. If we then consider, after BS$_1$,  the case in which a single-photon is reflected to Alice and 2$n$-1 photons are directed to Bob, the quantum state shared by Alice and Bob is an entangled state of the form,
\beq &&\ket{\Phi}^{(2n)}_{AB} = 1/\sqrt{2} \nonumber \\
&& (\ab{1}{0}{n-1}{n} + \ab{0}{1}{n}{n-1}).~~~~~ 
\label{eq:entangled2n} \eeq
To carry out RSP at Bob, Alice measures the single-photon in the projection basis $\ket{\phi}_A$. This will then project the quantum state of $(2n-1)$ photons at Bob to become,
\begin{gather}
\ket{\psi}^{(2n-1)}_{B} = \alpha \ket{n_H,(n-1)_V}_B+\beta e^{i\theta} \ket{(n-1)_H,n_V}_B,
\label{eq:2n-1photons}
\end{gather}
where the amplitudes $\alpha$ and $\beta$ as well as the phase $\theta$ are determined by the single-photon measurement at Alice.
To prepare a mixed state, similarly to above, it is necessary to introduce partial projection measurement $\mathcal{P}(\ket{\phi}_A\bra{\phi},p)$ in \eq{eq:partialpol}.  This then results in the multi-photon state of the form,
\beq \rho^{(2n-1)}_B &=& p \ket{\psi}^{(2n-1)}_B\bra{\psi} + (1-p) I^{(2n-1)}_B /2,
\label{eq:2n-1mixed} \eeq
where
\beq
I^{(2n-1)}_B &=& \ket{n_H,(n-1)_V}_B\bra{n_H,(n-1)_V} \nonumber \\ &+& \ket{(n-1)_H,n_V}_B\bra{(n-1)_H,n_V}.
\eeq
The RSP by single-photon measurement can, therefore, be applied to multi-photon states, which also provides enhanced engineering capability on multi-photon states.


In summary, we report the first experimental demonstration of remote preparation of three-photon entangled states by measuring only a single-photon entangled with the three photons. We have also generalized the RSP protocol to prepare a multi-photon entangled state with arbitrary photon number and purity via single-photon measurement. In addition to fundamental interest, our RSP protocol extends the capability of multi-photon state engineering beyond the linear optical limit via the nonlinearity induced by the single-photon measurement~\cite{Knill:2001aa,Lanyon:2008ci}, allowing us to prepare various multi-photon states required for quantum metrology ~\cite{Huver:2008aa,Demkowicz-Dobrzanski:2009aa,Matthews:2011aa,Tichy:2014aa} and for fundamental studies in quantum optics ~\cite{Sehat:2005ig, Bjork:2012aa}. We further anticipate that our work will stimulate fundamental studies on entanglement between a single particle and multiple particles, such as, nonlocality tests between a single-particle state and a multiple-particle state, quantum teleportation of a single-particle state to a multi-particle state, etc.


This work was supported by the National Research Foundation of Korea (Grant No. 2013R1A2A1A01006029). We thank S.-W. Lee for helpful discussions.

\section{Methods}

\textbf{Experimental setup.}
A femtosecond pulse laser (duration of 100 fs and average power of 165 mW) impinging on a BBO ($\beta$-BaB$_2$O$_4$, thickness of 2 mm) crystal generates two horizontally polarized photons at each of signal $s$ and idler $i$ modes, see Fig.~\ref{fig:setup}. To eliminate spectral and spatial correlations of the photons, interference filters (IFs, full width at half maximum bandwidth of 3 nm centered at 780 nm) and single-mode fibers (SMFs) are used. By using a half-wave plate H$_1$ at the angle of $\pi/4$, the polarization state of idler photons is rotated to vertical polarization just prior to the polarizing beam splitter (PBS$_1$). All the photons made to arrive at PBS$_1$ simultaneously and the quantum state of the photons at the output of PBS$_1$ is $\ket{2_H,2_V}$. Alice measures the single-photon reflected at the beam splitter (BS$_1$) by the single-photon detector (D$_1$): the measurement basis is controlled by a phase shifter (PS$_1(\theta)$) and a half wave plate (H$_2(\gamma)$). PBS$_2$ is used for RSP of a pure three-photon entangled state (\eq{eq:three-photon}) at Bob. RSP of a three-photon mixed state (\eq{eq:mixed}) requires the use of the partial polarizer (PP)~\cite{Peters:2005aa}. Bob measures the three photons by coincidence detection on D$_2$, D$_3$, and D$_4$. The measurement basis is controlled by using PS$_2(\phi)$ and H$_3(\delta)$. The phase shifters PS$_1(\theta)$ and PS$_2(\phi)$ are implemented by rotating a half-wave plate between two quarter wave plates set at the angle of $\pi/4$.

\textbf{CHSH inequality test.}
We obtain the CHSH parameter in \eq{eq:CHSH} by measuring the correlations values from joint measurements on the single-photon $\hat{\lambda}^\single$ and on the three photons $\hat{\lambda}^\three$:
\beq
\expec{\hat{\lambda}^\single}{\hat{\lambda}^\three} = \frac{ N_{+ +} + N_{- -} - N_{+ -} - N_{- +} } { N_{+ +} + N_{- -} + N_{+ -} + N_{- +} },
\eeq
where $N_{ij}$ is the frequency of outcomes $i\in(+,-)$ by $\hat{\lambda}^\single$ and $j\in(+,-)$ by $\hat{\lambda}^\three$.
For single-photon measurement, PS$_1$ is removed and PBS$_2$ is used. The outcomes $+$ and $-$ of $\hat{\lambda}^\single=\hat{\mu}^\single$ correspond to a click on D$_1$ when the angle $\gamma$ of H$_2$ is $\pi/16$ and $5\pi/16$, respectively. Similarly, $+$ and $-$ outcomes of $\hat{\lambda}^\single=\hat{\pi}^\single$ are obtained by setting the angle $\gamma$ of H$_2$ with $3\pi/16$ and $7\pi/16$, respectively. For three-photon measurement, PS$_2$ is removed, and coincidence clicks on D$_2$, D$_3$, and D$_4$ are recorded: the outcomes $(+,-)$ are obtained by setting the angle $\delta$ of H$_3$ at $(0, \pi/4)$ for $\hat{\lambda}^\three=\hat{\mu}^\three$ and at $(\pi/8, 3\pi/8)$ for $\hat{\lambda}^\three=\hat{\pi}^\three$.

\end{document}